\documentclass[12pt,a4paper]{article}
\usepackage{amsmath,amsfonts,amsthm,graphicx,lscape}
\usepackage[dvips]{psfrag}
\usepackage{cite}

\usepackage[normalem]{ulem}

\usepackage{textcomp}
\usepackage{amsmath,amsthm,amssymb,mathrsfs,cases}
\usepackage{amsfonts,graphicx,lscape}

\numberwithin{equation}{section}

\def\beqa{\begin{eqnarray}}
\def\enqa{\end{eqnarray}}
\def\beq{\begin{equation}}
\def\enq{\end{equation}}

\begin{document}
\title{
On 
a new integrable 
discretization
of the 
derivative nonlinear Schr\"odinger 
(Chen--Lee--Liu) 
equation 
}
\author{Takayuki \textsc{Tsuchida}
}
\maketitle
\begin{abstract} 
We 
propose 
a 
general 
integrable 
lattice system
involving 
some free parameters, 
which 
contains 
known 
integrable 
lattice 
systems such as 
the Ablowitz--Ladik 
discretization of the nonlinear Schr\"odinger (NLS) equation 
as 
special 
cases. 
With a suitable choice of the parameters, 
it provides 
a new 
integrable space-discretization 
of the 
derivative 
NLS equation
known as the Chen--Lee--Liu equation. 
Analogously to the continuous case, 
the space-discrete Chen--Lee--Liu system 
possesses a Lax pair 
and admits 
a complex conjugation reduction between the two dependent 
variables.  
Thus, 
we obtain a proper 
space-discretization of the 
Chen--Lee--Liu equation 
defined 
on the three 
lattice sites \mbox{$n-1$}, $n$, \mbox{$n+1$} 
for the first time. 
Considering 
a negative flow of 
the discrete Chen--Lee--Liu
hierarchy, 
we 
obtain 
a proper 
discretization of 
the massive Thirring model in light-cone coordinates. 
Multicomponent generalizations of the obtained 
discrete equations 
are straightforward because 
the performed 
computations are 
valid 
for the general case where 
the dependent variables are vector- or matrix-valued. 
\end{abstract}
%

\newpage
\section{Introduction}

There exists 
no 
systematic 
method 
of 
obtaining 
a proper discretization 
for 
a 
given 
integrable partial differential equation (PDE). 
For 
some 
scalar PDEs  
such as the sine-Gordon equation \mbox{$u_{xt}= \sin u$}\cite{Hiro77,Orfa1,Orfa2}, 
elementary 
auto-B\"acklund transformations
and the associated 
nonlinear superposition 
formula 
based on 
Bianchi's permutability theorem
can 
provide 
proper discretizations 
of the original 
continuous equations. 
However, 
this 
idea 
does not 
apply 
directly 
to 
complex-valued PDEs 
involving the operation of 
complex conjugation,  
such as 
the nonlinear Schr\"odinger (NLS)
equation \mbox{${\mathrm i} q_{t} + q_{xx} \pm 
2 
q^2 q^\ast = 0$}, 
which can  
be 
naturally 
obtained 
from 
two coupled PDEs 
by imposing 
a complex conjugation reduction
on the two dependent variables~\cite{AKNS73}. 
This is 
because 
elementary 
auto-B\"acklund transformations 
do not, 
in general, 
maintain the 
complex conjugation reduction~\cite{Kono82,Chud1,Chud2,DJM83}.  
One can 
consider 
more elaborate auto-B\"acklund transformations 
by composing 
two or more 
elementary 
auto-B\"acklund transformations 
so that  
the 
net result 
can maintain the complex conjugation 
reduction 
between 
the two dependent variables. 
However, 
such 
composite 
auto-B\"acklund transformations 
generally 
involve 
an indefinite 
integral~\cite{Calo76}
or 
a square-root function with 
an 
indefinite sign~\cite{Chen1,Lamb74} 
(also see (7.18) in~\cite{QNCL} 
or 
(4.17) in~\cite{Linden2}), 
so
they 
do not  
directly 
provide 
proper 
discretizations of the original PDEs, 
which 
can be written in 
local form and 
define 
the time evolution 
uniquely 
in the discrete setting. 

Because 
proper discretizations of integrable PDEs 
involving the operation of complex conjugation 
cannot be derived 
in a systematic manner, 
it is 
more 
productive and 
instructive 
to 
obtain 
such discretizations 
on a case-by-case 
consideration. 
In this paper, we 
propose a new proper space-discretization of 
an integrable 
derivative NLS equation
\mbox{${\mathrm i} q_{t} + q_{xx} + 
\mathrm{i} q q^\ast q_x= 0$},
usually 
referred to as the Chen--Lee--Liu equation~\cite{CLL} (also see~\cite{Chen80}). 
Actually, we already obtained an integrable space-discretization of 
the Chen--Lee--Liu equation in our previous paper~\cite{Tsuchi02}, but 
it depends on five lattice sites 
and 
is 
rather complicated. 
For the 
Chen--Lee--Liu 
system 
in the nonreduced 
form, 
i.e., two coupled PDEs 
for 
$q$ and $q^\ast$
wherein 
$q$ and $q^\ast$ are 
independent functions 
not related 
by a complex conjugation, 
a fully discrete analog 
was 
derived in~\cite{DJM83}  
(see also the 
continuous-time 
flows 
in~\cite{SY91,AY94,ASY00,Tsuchi02,TsuJMP11}); 
a correspondence between a fully discrete system and 
continuous-time flows is briefly 
described 
in~\cite{Comment}. 

We can obtain 
a proper 
space-discrete Chen--Lee--Liu equation 
defined on three lattice sites \mbox{$n-1$}, $n$, \mbox{$n+1$}
by constructing 
its 
Lax-pair representation. 
As the spatial part of the Lax-pair representation, 
we 
consider 
a new 
discrete spectral problem
implied by 
a binary 
B\"acklund--Darboux
transformation for the continuous Chen--Lee--Liu equation 
or, equivalently, that for 
the Ablowitz--Ladik 
lattice 
(an integrable discrete NLS equation~\cite{AL1,AL76}) 
proposed in \cite{Li92,Zullo}. 
Note that 
the Chen--Lee--Liu hierarchy and 
the Ablowitz--Ladik hierarchy are 
different facets of the same 
object 
and in a sense equivalent~\cite{Vek98,Vek02} 
(also see~\cite{Getmanov87,Getmanov93}). 
In contrast to the usual formulation of the 
B\"acklund--Darboux 
transformations~\cite{Kono82,Chud1,Chud2,
Sall82,MaSa91,Pemp}, 
we 
do not 
express 
two 
unknown functions 
appearing in 
the 
Darboux matrix 
explicitly in terms of 
linear eigenfunctions 
of the original spectral problem. 
Rather, we 
consider 
them as 
new 
dependent variables 
in the 
discrete spectral problem, 
which 
can 
be 
related to each other 
through 
a complex conjugation reduction. 
%
Then, 
we 
associate 
the 
discrete 
spectral problem 
with 
a suitable 
time-evolution equation, 
which comprises 
the Lax-pair representation; 
the compatibility condition 
provides 
a 
rather general 
lattice system involving 
some 
arbitrary parameters. 
Depending on the choice of the parameters, 
it 
provides 
a proper space-discretization of the Chen--Lee--Liu equation, 
as well as 
other 
integrable lattice 
systems such as the Ablowitz--Ladik lattice~\cite{AL1,AL76}. 

This paper is organized as follows.
In section 2, we introduce 
a discrete spectral problem and 
associate it with 
a suitable isospectral 
time-evolution equation. 
Then, 
the compatibility condition for this Lax-pair representation 
provides 
a new 
integrable lattice system; 
with a suitable choice of the parameters therein, 
we obtain a 
proper space-discretization of 
the 
Chen--Lee--Liu equation. 
In section 3, we change the time part of the Lax-pair 
representation 
to obtain 
a proper 
discretization of 
the massive Thirring model 
in light-cone (or characteristic) coordinates~\cite{KN2,Morris79,GIK}. 
Section 4 is devoted to concluding remarks. 
Throughout 
the paper, 
we perform 
the computations 
for 
the most 
general case where 
the dependent variables 
are 
(rectangular) matrix-valued, 
so 
multicomponent generalizations 
of the 
obtained 
equations 
are straightforward. 

\section{Space-discrete 
Chen--Lee--Liu equation}

In this section, we propose 
a new integrable lattice system, 
which contains 
a 
proper 
space-discretization of the 
Chen--Lee--Liu equation as a special case, 
through a Lax-pair 
representation. 
In subsection~\ref{sub2.1}, we introduce 
a new discrete 
spectral problem, 
which gives 
the spatial part of the Lax pair. 
In subsection~\ref{sub2.2}, 
we associate it with 
an isospectral 
time-evolution equation, 
which is 
the temporal part of the Lax pair. 

\subsection{B\"acklund--Darboux transformation 
as a discrete spectral problem}
\label{sub2.1}

The 
Ablowitz--Ladik spectral problem~\cite{AL1,AL76} 
can be generalized to 
a block-matrix form~\cite{GI82} 
(also see~\cite{Tsuchi02,DM2010} and references therein)
as given by 
\begin{align}
& 
\left[
\begin{array}{c}
 \Psi_{1,m+1} \\
 \Psi_{2,m+1} \\
\end{array}
\right]
=\left[
\begin{array}{cc}
 \zeta I & Q_m \\
 R_m & \frac{1}{\zeta}I \\
\end{array}
\right]
\left[
\begin{array}{c}
 \Psi_{1,m} \\
 \Psi_{2,m} \\
\end{array}
\right]. 
\label{AL-Ln}
\end{align}
Here, $\zeta$ is a constant 
spectral parameter and 
$Q_m$ and $R_m$ are, 
respectively,  
\mbox{$l_1 \times l_2$}
and \mbox{$l_2 \times l_1$} (generally rectangular) 
matrices. 
Thus, 
the square matrix above 
is 
partitioned 
as 
an \mbox{$(l_1+l_2) \times (l_1+l_2)$} block matrix; 
for notational brevity, 
we 
omit the index of each 
unit 
matrix $I$ to indicate its 
size. 

Among 
an 
infinite 
set 
of isospectral time-evolution equations 
compatible with the 
Ablowitz--Ladik spectral 
problem (\ref{AL-Ln}) (cf.~\cite{Chiu77,Kako}), 
the 
most 
fundamental
ones 
are 
\begin{align}
& \left[
\begin{array}{c}
 \Psi_{1,m} \\
 \Psi_{2,m} \\
\end{array}
\right]_{t_1}
= 
\left[
\begin{array}{cc}
 \zeta^2 I -Q_m R_{m-1} & \zeta  Q_m \\
 \zeta R_{m-1} & O \\
\end{array}
\right]
\left[
\begin{array}{c}
 \Psi_{1,m} \\
 \Psi_{2,m} \\
\end{array}
\right]
\label{AL-Mn1}
\end{align}
and
\begin{align}
& \left[
\begin{array}{c}
 \Psi_{1,m} \\
 \Psi_{2,m} \\
\end{array}
\right]_{t_{-1}}
= 
\left[
\begin{array}{cc}
 O & \frac{1}{\zeta} Q_{m-1}\\
 \frac{1}{\zeta} R_{m} & 
 \frac{1}{\zeta^2} I - R_m Q_{m-1} \\
\end{array}
\right]
\left[
\begin{array}{c}
 \Psi_{1,m} \\
 \Psi_{2,m} \\
\end{array}
\right].
\label{AL-Mn2}
\end{align}
Here, the subscript \mbox{$t_j
$} 
denotes 
time 
differentiation 
and the symbol italic $O$ 
represents 
a zero matrix. 
The corresponding 
equations of motion
are derived from 
the compatibility condition 
for the overdetermined linear system, 
(\ref{AL-Ln}) and (\ref{AL-Mn1}) or (\ref{AL-Mn2}), i.e., 
\begin{equation} 
\label{sd-AL1}
\left\{ 
\begin{split}
& Q_{m, t_1}- Q_{m+1} + Q_{m+1} R_m Q_m =O, 
\\[0.5mm]
& R_{m, t_1}+ R_{m-1} - R_m Q_m R_{m-1} =O,  
\end{split} 
\right. 
\end{equation}
and
\begin{equation} 
\label{sd-AL2}
\left\{ 
\begin{split}
& Q_{m,t_{-1}}+ Q_{m-1} - Q_m R_m Q_{m-1} =O, 
\\[0.5mm]
& R_{m,t_{-1}}- R_{m+1} + R_{m+1} Q_m R_m =O. 
\end{split} 
\right. 
\end{equation}
These are the two elementary flows of the Ablowitz--Ladik hierarchy
and 
a suitable linear combination of them together with 
the trivial 
zeroth flow 
provides 
the 
Ablowitz--Ladik 
discretization of the NLS system~\cite{AL1,Suris97'}. 
Note that the use of $O$, instead of $0$, 
on the right-hand side of the equations
implies that the dependent variables 
are matrix-valued. 

A crucial 
fact 
noticed 
by Vekslerchik~\cite{Vek98,Vek02} 
(also see Barashenkov--Getmanov~\cite{Getmanov87,Getmanov93}) 
is that 
the Ablowitz--Ladik hierarchy 
is in a sense equivalent to 
a derivative NLS hierarchy 
called 
the Chen--Lee--Liu hierarchy~\cite{CLL}. 
This is clear 
from the observation that the isospectral equation 
(\ref{AL-Mn1}) with \mbox{$t_{1} \to x$} 
(or (\ref{AL-Mn2}) with \mbox{$t_{-1} \to x$}) 
has 
essentially 
the same form as the spectral problem associated with 
the continuous Chen--Lee--Liu hierarchy~\cite{WS,Dodd83,Dodd84} 
(see~\cite{Linden1,TW3} for the matrix case);
note that 
this spectral problem can be rewritten in a traceless form 
using 
a simple gauge transformation~\cite{Getmanov87}. 
Actually, 
the Chen--Lee--Liu spectral problem was 
first presented 
for 
the massive Thirring model in light-cone (or characteristic) 
coordinates~\cite{KN2,Morris79,GIK} 
(also see~\cite{Kuz,KaMoIno} for the case of laboratory coordinates),  
which is the first negative flow of the Chen--Lee--Liu hierarchy~\cite{NCQL,Linden1}; 
the compatibility condition 
for the overdetermined linear system, 
(\ref{AL-Mn1}) and (\ref{AL-Mn2}) 
for any fixed value of $m$, 
indeed provides 
the equations of motion for 
the massive Thirring model 
((\ref{sd-AL1}) for $Q_{m-1,t_1}$ and $R_{m,t_1}$ 
and (\ref{sd-AL2}) for $Q_{m,t_{-1}}$ and $R_{m-1,t_{-1}}$). 
Conversely, we can regard 
the Ablowitz--Ladik spectral problem (\ref{AL-Ln}) 
as a B\"acklund--Darboux
transformation for the continuous Chen--Lee--Liu hierarchy~\cite{Getmanov87,Getmanov93}; 
then 
the elementary flow 
(\ref{sd-AL1}) (or (\ref{sd-AL2})) of the Ablowitz--Ladik hierarchy 
can be considered 
as an 
auto-B\"acklund transformation 
for the continuous Chen--Lee--Liu hierarchy. 
These items of information 
imply 
that a 
suitably chosen 
B\"acklund--Darboux transformation 
for the Ablowitz--Ladik hierarchy 
can 
provide 
a proper discretization 
of the Chen--Lee--Liu 
spectral problem. 

In fact, 
B\"acklund--Darboux transformations 
for the Ablowitz--Ladik hierarchy 
have 
already 
been 
considered 
in 
a number of papers (see, {\em e.g.}, \cite{Li92,Pemp,NQC,Rourke}); 
those 
results 
are closely related 
to 
integrable 
time-discretizations of 
the 
flows of 
the 
Ablowitz--Ladik 
hierarchy~\cite{Suris97',2010JPA,Chiu77,Zullo,AL77,Vek02}.  
Noting that the Ablowitz--Ladik hierarchy is 
invariant under the rescaling 
\mbox{$Q_m \to k Q_m$}, \mbox{$R_m \to k^{-1} R_m$}, 
we 
can  
introduce 
a slightly generalized 
version 
of 
the binary B\"acklund--Darboux transformation 
in the form (cf.~\cite{Li92,Zullo,Adler08}): 
\begin{align}
& \left[
\begin{array}{c}
 \widetilde{\Psi}_{1,m}  \\
 \widetilde{\Psi}_{2,m} \\
\end{array}
\right]
= 
\left\{ \left[
\begin{array}{cc}
 \left( \alpha \zeta + \frac{\delta}{\zeta} \right) I &  \\
  & \left( \gamma \zeta + \frac{\beta}{\zeta} \right) I \\
\end{array}
\right] + \left( \alpha \beta - \gamma \delta \right)
 \left[
\begin{array}{cc}
\gamma \zeta I & u_m \\
 v_m & \frac{\delta}{\zeta} I \\
\end{array}
\right]^{-1}
\right\}
\left[
\begin{array}{c}
 \Psi_{1,m}  \\
 \Psi_{2,m} \\
\end{array}
\right]. 
\label{ALDB1}
\end{align}
Here, $\alpha$, $\beta$, $\gamma$ and $\delta$ are arbitrary parameters 
and $u_m$ and $v_m$ are 
intermediate potentials
that can be expressed in terms of 
linear eigenfunctions of the Ablowitz--Ladik 
spectral problem (\ref{AL-Ln}) at special values of the spectral parameter 
$\zeta$; 
however, this information 
is not necessary 
for 
the purpose of 
this paper and 
we will treat $u_m$ and $v_m$ as 
new independent unknowns. 
Alternatively, we can consider 
(\ref{ALDB1})
as a generalized binary B\"acklund--Darboux transformation for 
the isospectral problem (\ref{AL-Mn1}) 
or (\ref{AL-Mn2}) 
associated with 
the Chen--Lee--Liu hierarchy. 
The binary B\"acklund--Darboux transformation 
for 
the Chen--Lee--Liu hierarchy 
(or, more specifically, 
the massive Thirring model) 
was apparently first considered 
by 
P.~I.~Holod (also written as Golod, preprint in Russian, Kiev, 
1978) in a somewhat preliminary form; 
the result can be found in more accessible papers~\cite{Morris80, Pri81, 
David84} 
(see~\cite{Getmanov93,DHS84,Getmanov87,BGK93} for further developments). 

Because the Ablowitz--Ladik hierarchy is invariant 
under a space translation \mbox{$m \to m+l$}, 
B\"acklund--Darboux transformations 
can always be 
combined with a space translation; 
for instance, 
using (\ref{ALDB1}) and (\ref{AL-Ln}), 
we can compute 
\[
\left[
\begin{array}{c}
 \widetilde{\Psi}_{1,m+1}  \\
 \widetilde{\Psi}_{2,m+1} \\
\end{array}
\right] 
\;\, \mathrm{or} \;\, 
\left[
\begin{array}{c}
 \widetilde{\Psi}_{1,m-1}  \\
 \widetilde{\Psi}_{2,m-1} \\
\end{array}
\right]
\]
and consider it as 
a new B\"acklund--Darboux transformation, 
which has 
the same $\zeta$-dependence as that 
usually employed in 
the time-discretizations of 
the Ablowitz--Ladik flows~\cite{AL76,AL77,Suris97',2010JPA}. 
Note also that an overall factor of (\ref{ALDB1}) is nonessential, 
so we can multiply the right-hand side of (\ref{ALDB1}) by 
$\zeta$ or $1/\zeta$ to 
obtain 
more familiar $\zeta$-dependence 
that appears in the 
time-discretizations of the 
Ablowitz--Ladik flows. 

Now, we freeze the lattice index $m$ 
for the Ablowitz--Ladik hierarchy 
and 
reinterpret the B\"acklund--Darboux transformation (\ref{ALDB1}) 
as defining 
a new discrete spectral problem:  
\begin{equation}
\left[
\begin{array}{c}
 \Psi_{1, n+1}  \\
 \Psi_{2, n+1} \\
\end{array}
\right]
= L_n 
\left[
\begin{array}{c}
 \Psi_{1,n}  \\
 \Psi_{2,n} \\
\end{array}
\right], 
\label{gDB2}
\end{equation}
%
where 
$n$ can be intuitively 
understood as the number of iterations 
of the B\"acklund--Darboux transformation. 
Here, 
the 
Lax matrix $L_n$ 
is given 
by 
\begin{subequations}
\label{Lax-Ln1}
\begin{align}
L_n & =   \left[
\begin{array}{cc}
 \left( \alpha \zeta + \frac{\delta}{\zeta} \right) I &  \\
  & \left( \gamma \zeta + \frac{\beta}{\zeta} \right) I \\
\end{array}
\right] + \left( \alpha \beta - \gamma \delta \right)
 \left[
\begin{array}{cc}
\gamma \zeta I & u_n \\
 v_n & \frac{\delta}{\zeta} I \\
\end{array}
\right]^{-1}
\label{1st_exp}
\\[2mm]
& =  \left[
\begin{array}{cc}
 \left( \alpha \zeta + \frac{\delta}{\zeta} \right) I &  \\
  & \left( \gamma \zeta + \frac{\beta}{\zeta} \right) I \\
\end{array}
\right] 
\nonumber \\[1mm]
& \hphantom{=} \;\, \mbox{}
 + \left( \alpha \beta - \gamma \delta \right)
 \left[
\begin{array}{cc}
\frac{\delta}{\zeta} (\gamma \delta I - u_n v_n)^{-1}  
 & -(\gamma \delta I - u_n v_n)^{-1}u_n \\
 -(\gamma \delta I - v_n u_n)^{-1}v_n 
 & \gamma \zeta (\gamma \delta I - v_n u_n)^{-1}  \\
\end{array}
\right]
\label{2nd_exp}
\\[2mm]
&= \left[
\begin{array}{cc}
\alpha \zeta I & u_n \\
 v_n &  \frac{\beta}{\zeta} I \\
\end{array}
\right]
\left[
\begin{array}{cc}
\left( \gamma\zeta + \frac{\beta}{\zeta} \right)  I &  \\
  &  \left( \alpha \zeta + \frac{\delta}{\zeta} \right) I \\
\end{array}
\right] 
 \left[
\begin{array}{cc}
\gamma \zeta I & u_n \\
 v_n & \frac{\delta}{\zeta} I \\
\end{array}
\right]^{-1}
\label{3rd_exp}
\\[2mm]
&= \left[
\begin{array}{cc}
\gamma \zeta I & u_n \\
 v_n & \frac{\delta}{\zeta} I \\
\end{array}
\right]^{-1} 
\left[
\begin{array}{cc}
\left( \gamma\zeta + \frac{\beta}{\zeta} \right)  I &  \\
  &  \left( \alpha \zeta + \frac{\delta}{\zeta} \right) I \\
\end{array}
\right] 
\left[
\begin{array}{cc}
\alpha \zeta I & u_n \\
 v_n &  \frac{\beta}{\zeta} I \\
\end{array}
\right]. 
\label{4th_exp}
\end{align}
\end{subequations}
The 
parameters 
$\alpha$, $\beta$, $\gamma$ and $\delta$ 
can 
be varied 
at each 
application of 
the B\"acklund--Darboux transformation, 
so they 
can 
be 
arbitrary functions of 
the 
discrete independent variable $n$; 
however, 
for simplicity 
we 
consider 
them as constants. 
Each 
of the 
four equivalent 
expressions 
in (\ref{Lax-Ln1}) 
has its own advantages. 
%

\subsection{Isospectral 
time-evolution equation}
\label{sub2.2}

To 
compose 
a Lax pair, 
we 
associate 
(\ref{gDB2}) with 
a suitable 
isospectral 
time-evolution equation, 
\begin{equation}
\left[
\begin{array}{c}
 \Psi_{1, n}  \\
 \Psi_{2, n} \\
\end{array}
\right]_t
= M_n 
\left[
\begin{array}{c}
 \Psi_{1,n}  \\
 \Psi_{2,n} \\
\end{array}
\right]. 
\label{general_M}
\end{equation}
%
The compatibility condition 
for the overdetermined linear system, 
(\ref{gDB2}) and (\ref{general_M}), 
is given by (a space-discrete 
version of) the zero-curvature 
equation~\cite{AL1,Kako,AL76,
Ize81}
\begin{equation}
 L_{n,t} +L_n M_n - M_{n+1}L_n = O, 
\label{Lax_eq}
\end{equation}
where
$L_n$ and $M_n$  
comprise 
the 
Lax pair. 
For the Lax matrix $L_n$ 
in (\ref{Lax-Ln1}), 
the first expression (\ref{1st_exp})
allows us to compute $L_{n,t}$
in such a way that 
the 
time-evolution 
equations 
for $u_{n}$ and $v_{n}$
can be 
obtained 
from (\ref{Lax_eq})
explicitly; 
note that \mbox{$(X^{-1})_t = - X^{-1} X_t X^{-1}$} 
for a square matrix $X$.

To 
obtain 
a proper 
space-discrete analog  
of 
the 
Chen--Lee--Liu system, 
we 
consider a temporal Lax matrix $M_n$
that 
goes well 
with 
the factorized form (\ref{3rd_exp}) or (\ref{4th_exp}) 
of the spatial 
Lax matrix $L_n$.  
Thus, a
natural 
ansatz for $M_n$
is 
\begin{align}
& M_n 
= \left( \alpha \beta - \gamma \delta \right) 
\left[
\begin{array}{cc}
 \frac{\beta}{\zeta} I & -u_n \\
 -v_n &  \alpha \zeta I \\
\end{array}
\right]
\left[
\begin{array}{cc}
\frac{\gamma \zeta}{\gamma \zeta + \frac{\beta}{\zeta}} F_n & \\
  & \frac{\frac{\delta}{\zeta}}{\alpha \zeta + \frac{\delta}{\zeta} } G_n \\
\end{array}
\right] 
\left[
\begin{array}{cc}
 \frac{\beta}{\zeta} I & -u_{n-1} \\
 -v_{n-1} &  \alpha \zeta  I \\
\end{array}
\right]
+ 
\left[
\begin{array}{cc}
c I & \\
 & d I \\
\end{array}
\right],  
\label{Lax-Mn1}
\end{align}
where $F_n$ and $G_n$ 
are 
square matrices, 
and $c$ and $d$ are 
arbitrary 
constants 
introduced for adding 
the trivial zeroth flow. 
The condition \mbox{$\alpha \beta - \gamma \delta \neq 0$} 
is assumed 
for the Lax pair to be nontrivial. 
Substituting (\ref{Lax-Ln1}) and (\ref{Lax-Mn1}) 
into (\ref{Lax_eq}), 
we 
obtain 
recursion relations for determining 
$F_n$ and $G_n$; 
they are 
satisfied 
by setting 
\begin{align}
F_n = a 
\left( \beta \gamma I - u_{n-1} v_n \right)^{-1} 
,
\hspace{5mm}
G_n = b 
\left( \alpha \delta I - v_{n-1} u_n \right)^{-1} 
, 
\label{FG_choice}
\end{align}
where 
$a$ and $b$ are arbitrary constants. 
We can also consider the more general case where 
$a$ and $b$, as well as $c$ and $d$ in (\ref{Lax-Mn1}), 
are arbitrary functions of 
time $t$, but 
we do not 
discuss 
it 
in this paper. 
Then, 
(\ref{Lax_eq}) 
for 
(\ref{Lax-Ln1}) and (\ref{Lax-Mn1}) 
with (\ref{FG_choice})
provides 
an evolutionary lattice system: 
\begin{equation} 
\label{sdCLL1}
\left\{ 
\begin{split}
& u_{n,t}  - a \gamma \left( \alpha \beta I - u_{n} v_n \right)
	\left( \beta \gamma I - u_{n-1} v_n \right)^{-1} 
	\left( \beta u_n - \delta u_{n-1} \right)
\\ 
& 
\mbox{} - b \delta \left( \gamma u_{n+1} -\alpha u_{n} \right)
 \left( \alpha \delta I - v_{n} u_{n+1} \right)^{-1} 
 \left( \alpha \beta I - v_{n} u_n \right)
 - (c-d) u_n = O, 
\\[1.5mm]
& v_{n,t}  - b \delta \left( \alpha \beta I - v_{n} u_n \right)
	\left( \alpha \delta I - v_{n-1} u_n \right)^{-1} 
	\left( \alpha v_n - \gamma v_{n-1} \right)
\\ 
& \mbox{} - a \gamma \left( \delta v_{n+1} - \beta v_n \right) 
 \left( \beta \gamma I - u_{n} v_{n+1} \right)^{-1} 
 \left( \alpha \beta I - u_{n} v_n \right)
 + (c-d) v_n = O. 
\end{split}
\right.
\end{equation}
In the case of \mbox{$\beta=\alpha^\ast$}, \mbox{$\delta=\gamma^\ast$}, 
\mbox{$b=a^\ast$} and \mbox{$d=c^\ast$}, 
system (\ref{sdCLL1}) admits the complex conjugation reduction 
\mbox{$v_n=\sigma u_n^\ast$} with a real constant $\sigma$. 
To consider a Hermitian conjugation reduction between $u_n$ and $v_n$, 
we need to rewrite (\ref{sdCLL1}).  
By setting 
\mbox{$c-d = -a \left( \alpha \beta - \gamma \delta \right)$} and 
using 
the identities 
\begin{align}
& \gamma \left( \alpha \beta I - u_{n} v_n \right)
	\left( \beta \gamma I - u_{n-1} v_n \right)^{-1} 
	\left( \beta u_n - \delta u_{n-1} \right) 
	 - \left( \alpha \beta - \gamma \delta \right) u_n 
\nonumber \\
&= \delta \left( \gamma u_n - \alpha u_{n-1} \right)  
 - \left( \gamma u_n - \alpha u_{n-1} \right) v_n
 \left( \beta \gamma I - u_{n-1} v_n \right)^{-1} 
 \left( \beta u_n - \delta u_{n-1} \right)
\nonumber \\
&= \beta \left( \gamma u_n - \alpha u_{n-1} \right) 
 \left( \beta \gamma I - v_n u_{n-1} \right)^{-1} 
 \left( \gamma \delta I - v_{n} u_n \right), 
\nonumber 
\\[3mm]
%
& \gamma \left( \delta v_{n+1} - \beta v_n \right) 
 \left( \beta \gamma I - u_{n} v_{n+1} \right)^{-1} 
 \left( \alpha \beta I - u_{n} v_n \right) 
 + \left( \alpha \beta - \gamma \delta \right) v_n
\nonumber \\
&= \delta  \left( \alpha v_{n+1} - \gamma v_n \right) 
+ \left( \delta v_{n+1} - \beta v_n \right) 
 \left( \beta \gamma I - u_{n} v_{n+1} \right)^{-1} u_{n} 
 \left( \alpha v_{n+1} - \gamma v_n \right) 
\nonumber \\
&= \beta \left( \gamma \delta I -  v_{n} u_{n}  \right) 
 \left( \beta \gamma I -  v_{n+1} u_{n} \right)^{-1}  
 \left( \alpha v_{n+1} - \gamma v_n \right), 
\nonumber 
\end{align}
we can rewrite (\ref{sdCLL1}) 
as 
\begin{equation} 
\label{sdCLL2}
\left\{ 
\begin{split}
 u_{n,t} &- a \beta \left( \gamma u_n - \alpha u_{n-1} \right) 
 \left( \beta \gamma I - v_n u_{n-1} \right)^{-1} 
 \left( \gamma \delta I - v_{n} u_n \right)
\\ 
& 
\mbox{} - b \delta \left( \gamma u_{n+1} -\alpha u_{n} \right)
 \left( \alpha \delta I - v_{n} u_{n+1} \right)^{-1} 
 \left( \alpha \beta I - v_{n} u_n \right) = O, 
\\[1.5mm]
 v_{n,t} &- b \delta \left( \alpha \beta I - v_{n} u_n \right)
	\left( \alpha \delta I - v_{n-1} u_n \right)^{-1} 
	\left( \alpha v_n - \gamma v_{n-1} \right)
\\ 
& \mbox{} - a \beta \left( \gamma \delta I -  v_{n} u_{n}  \right) 
 \left( \beta \gamma I -  v_{n+1} u_{n} \right)^{-1}  
 \left( \alpha v_{n+1} - \gamma v_n \right) = O. 
\end{split}
\right.
\end{equation}

System (\ref{sdCLL1}) (or (\ref{sdCLL2}))
is a 
rather 
general 
system 
involving 
several 
free parameters
and thus encompasses 
simpler 
lattice systems 
as 
particular 
(or limiting)  
cases; 
some of them are already 
known. 
\begin{itemize}
\item 
By rescaling the dependent variables and 
parameters 
as 
\mbox{$u_n v_n 
= \alpha \beta u'_n v'_n$}, \mbox{$a=a'/\alpha$} and \mbox{$b=b'/\beta$}, 
omitting the prime 
and 
taking the limit \mbox{$\alpha, \beta \to 0$}, 
(\ref{sdCLL1}) reduces to 
the matrix generalization~\cite{GI82} 
(also see~\cite{Tsuchi02,DM2010} and references therein)
of 
the Ablowitz--Ladik lattice~\cite{AL1}: 
\begin{equation} 
\nonumber 
\left\{ 
\begin{split}
& u_{n,t}  + a \delta \left( I - u_{n} v_n \right)  u_{n-1} 
- b  \gamma u_{n+1} 
 \left( I - v_{n} u_n \right)
 - (c-d) u_n = O, 
\\[0.5mm]
& v_{n,t}  + b \gamma \left( I - v_{n} u_n \right) v_{n-1} 
- a \delta v_{n+1} \left( I - u_{n} v_n \right)
 + (c-d) v_n = O. 
\end{split}
\right.
\end{equation}
This is a linear combination of the 
two elementary flows 
(\ref{sd-AL1}) 
and (\ref{sd-AL2}) 
and the trivial zeroth flow
of the Ablowitz--Ladik hierarchy. 
\item
By rescaling the dependent variables
and 
parameter 
as 
\mbox{$u_n v_n 
= \beta \gamma u'_n v'_n$}
and \mbox{$b=b'/\beta$}, 
omitting the prime 
and 
taking the limit \mbox{$\beta \to 0$}, 
(\ref{sdCLL2}) reduces to 
the already known 
space-discretization of the Chen--Lee--Liu 
system~\cite{Tsuchi02,TsuJMP11}: 
\begin{equation} 
\nonumber 
\left\{ 
\begin{split}
 u_{n,t} &- a \delta \left( \gamma u_n - \alpha u_{n-1} \right) 
 \left( I - v_n u_{n-1} \right)^{-1} 
- \frac{b}{\alpha} \left( \gamma u_{n+1} -\alpha u_{n} \right)
 \left( \alpha I - \gamma v_{n} u_n \right) = O, 
\\
 v_{n,t} &- \frac{b}{\alpha} \left( \alpha I - \gamma v_{n} u_n \right)
	\left( \alpha v_n - \gamma v_{n-1} \right)
- a \delta 
 \left( I - v_{n+1} u_{n} \right)^{-1}  
 \left( \alpha v_{n+1} - \gamma v_n \right) = O. 
\end{split}
\right.
\end{equation}
In the case of 
scalar $u_n$ and $v_n$, 
this 
system was 
previously 
studied 
in~\cite{SY91,AY94,ASY00}; 
in fact, it 
can be 
considered as 
a continuous-time analog of the fully discrete Chen--Lee--Liu 
system proposed by Date, Jimbo and Miwa~\cite{DJM83}, 
associated with an elementary auto-B\"acklund 
transformation for the continuous Chen--Lee--Liu hierarchy 
(cf.~\cite{Comment}). 
Unfortunately, 
this system does not admit 
a complex/Hermitian conjugation reduction between 
$u_n$ and $v_n$, 
so it 
cannot 
provide 
a proper discretization 
of the Chen--Lee--Liu equation. 
%
\item
In the special case of \mbox{$\alpha=-\delta$} and \mbox{$\beta=-\gamma$}, 
(\ref{sdCLL1}) reduces to a 
nontrivial lattice system 
\begin{equation} 
\nonumber 
\left\{ 
\begin{split}
& u_{n,t}  - a \gamma \left( \gamma \delta I - u_{n} v_n \right)
	\left( \gamma^2 I + u_{n-1} v_n \right)^{-1} 
	\left( \gamma u_n + \delta u_{n-1} \right)
\\ 
& 
\mbox{} + b \delta \left( \gamma u_{n+1} +\delta u_{n} \right)
 \left( \delta^2 I + v_{n} u_{n+1} \right)^{-1} 
 \left( \gamma \delta I - v_{n} u_n \right)
 - (c-d) u_n = O, 
\\[1.5mm]
& v_{n,t}  - b \delta \left( \gamma \delta I - v_{n} u_n \right)
	\left( \delta^2 I + v_{n-1} u_n \right)^{-1} 
	\left( \delta v_n + \gamma v_{n-1} \right)
\\ 
& \mbox{} + a \gamma \left( \delta v_{n+1} +\gamma v_n \right) 
 \left( \gamma^2 I + u_{n} v_{n+1} \right)^{-1} 
 \left( \gamma \delta I - u_{n} v_n \right)
 + (c-d) v_n = O, 
\end{split}
\right.
\end{equation}
while 
the Lax pair
given by (\ref{Lax-Ln1}) and (\ref{Lax-Mn1})
becomes trivial 
in 
the naive 
limit 
\mbox{$\alpha \to -\delta$}, 
\mbox{$\beta \to -\gamma$}.  
This lattice system can be derived from 
a B\"acklund--Darboux transformation for the continuous NLS 
hierarchy in 
the same manner 
as (\ref{sdCLL1}) 
is derived from a B\"acklund--Darboux transformation 
for the Ablowitz--Ladik hierarchy. 
By further setting 
\mbox{$\gamma=\delta=\pm 1$}, we obtain 
a new proper space-discretization 
of the matrix NLS system 
as well as the 
matrix modified KdV system. 
We will discuss it in a separate paper. 
\\
\end{itemize}
%
Not all of the four 
parameters 
$\alpha$, $\beta$, $\gamma$ and $\delta$ in 
(\ref{sdCLL1}) (or (\ref{sdCLL2})) 
are essential; 
the number of independent parameters can be reduced 
by applying 
a point transformation of the form: 
\mbox{$u_n = \mu \nu^n u_n'$}, \mbox{$v_n = \nu^{-n} v_n'$} 
with nonzero constants $\mu$ and $\nu$. 
However, 
it then becomes difficult to 
consider 
limiting cases where 
one or more parameters tend to zero 
as mentioned 
above, so we prefer 
the 
``redundant" expression 
(\ref{sdCLL1}) (or (\ref{sdCLL2})).  

%
By setting \mbox{$\gamma=\alpha$},  
\mbox{$\delta=-\beta$} and \mbox{$\alpha \beta=1$}, 
(\ref{sdCLL2}) 
provides 
a new integrable 
space-discretization of the Chen--Lee--Liu system: 
\begin{equation} 
\label{sdCLL3}
\left\{ 
\begin{split}
 u_{n,t} &+ a \left( u_n - u_{n-1} \right) 
 \left( I - v_n u_{n-1} \right)^{-1} 
 \left( I + v_{n} u_n \right)
\\ 
& 
\mbox{} - b \left( u_{n+1} - u_{n} \right)
 \left( I + v_{n} u_{n+1} \right)^{-1} 
 \left( I - v_{n} u_n \right) = O, 
\\[1.5mm]
 v_{n,t} &- b \left( I - v_{n} u_n \right)
	\left( I + v_{n-1} u_n \right)^{-1} 
	\left( v_n -  v_{n-1} \right)
\\ 
& \mbox{} + a \left( I + v_{n} u_{n}  \right) 
 \left( I -  v_{n+1} u_{n} \right)^{-1}  
 \left( v_{n+1} -  v_n \right) = O. 
\end{split}
\right.
\end{equation}
System (\ref{sdCLL3}) with \mbox{$b=-a^\ast$} 
admits the Hermitian conjugation reduction 
\mbox{$v_n = \mathrm{i} u_n^\dagger$}; 
in the case of \mbox{$a=b=\mathrm{i}$},  
we obtain 
a
proper 
space-discretization of the Chen--Lee--Liu equation~\cite{CLL} 
(see~\cite{Linden1,Olver2,TW3,Ad,Dimakis} for the matrix case): 
\begin{align}
 \mathrm{i} u_{n,t} & + \left( u_{n+1} - u_{n} \right)
 \left( I + \mathrm{i} u_{n}^\dagger u_{n+1} \right)^{-1} 
 \left( I - \mathrm{i} u_{n}^\dagger u_n \right) 
\nonumber \\ 
& 
\mbox{} - \left( u_n - u_{n-1} \right) 
 \left( I - \mathrm{i} u_n^\dagger u_{n-1} \right)^{-1} 
 \left( I + \mathrm{i} u_n^\dagger u_n \right)= O. 
\nonumber
\end{align}
In the scalar case, this 
reads 
\begin{align}
 \mathrm{i} u_{n,t}  + 
 \frac{1 - \mathrm{i} |u_n|^2 }{1 + \mathrm{i} u_{n}^\ast u_{n+1}}
 \left( u_{n+1} - u_{n} \right)
- \frac{1 + \mathrm{i} |u_n|^2}{1 - \mathrm{i} u_n^\ast u_{n-1}}
 \left( u_n - u_{n-1} \right) =0, 
\nonumber 
\end{align}
or equivalently, 
\begin{align}
 \mathrm{i} u_{n,t}  + \left( u_{n+1}  + u_{n-1} - 2u_{n} \right)
 - \mathrm{i}  \frac{\left( u_{n+1}^2 - u_{n}^2 \right) u_{n}^\ast}
	{1 + \mathrm{i} u_{n}^\ast u_{n+1}}
- \mathrm{i}\frac{\left( u_n^2 - u_{n-1}^2 \right)u_n^\ast}
	{1 - \mathrm{i} u_n^\ast u_{n-1}}=0. 
\nonumber 
\end{align}
In addition, 
system (\ref{sdCLL3}) with \mbox{$b=-a$} 
admits the matrix transpose 
reduction \mbox{$v_n = u_n^T C$}, where $C$ is a constant skew-symmetric 
matrix~\cite{TsuJMP10}; in particular, in the case of 
\mbox{$a=1$} and \mbox{$b=-1$}, we obtain  
\begin{align}
 u_{n,t} & + \left( u_{n+1} - u_{n} \right)
 \left( I + u_n^T C u_{n+1} \right)^{-1} 
 \left( I - u_n^T C u_n \right) 
\nonumber \\ 
& 
\mbox{} +  \left( u_n - u_{n-1} \right) 
 \left( I - u_n^T C u_{n-1} \right)^{-1} 
 \left( I + u_n^T C u_n \right)
= O, \hspace{5mm} C^T = -C. 
\nonumber 
\end{align}

In the scalar case, 
(\ref{sdCLL1}) 
is a Hamiltonian system 
with 
an ultralocal 
(but 
noncanonical) Poisson structure. 
Indeed, 
it can be expressed 
as 
\mbox{$u_{n,t} = \{ u_n, H \}$} and \mbox{$v_{n,t} = \{ v_n, H \}$}, 
where the Hamiltonian and the Poisson brackets are given by 
\[
H = 
\sum_n \left[ a \log \left( \frac{ \beta \gamma - u_{n-1} v_{n}}
	{\gamma \delta - u_{n} v_{n}} \right)
 - b \log \left( \frac{ \alpha \delta - u_{n+1} v_{n}}
	{\gamma \delta - u_{n} v_{n}} \right) 
 + \frac{c-d}{\alpha \beta - \gamma \delta} \log \left( 
\frac{\alpha \beta - u_{n} v_{n}}{\gamma \delta - u_{n} v_{n}} \right) 
\right]
\]
and 
\[
\{ u_m, u_n \} = \{ v_m, v_n \} =0, \hspace{5mm} 
\{ u_m, v_n \} = \delta_{m n} \left( \alpha \beta - u_{n} v_{n} \right)
	\left( \gamma \delta - u_{n} v_{n} \right), 
\]
respectively. Here, $\delta_{m n}$ is 
the Kronecker delta, which should not be confused with the 
free 
parameter $\delta$. 
This Hamiltonian structure encompasses 
the already known 
Hamiltonian structures for the simpler lattice systems 
in the scalar case~\cite{Kako,Kulish,Wang13,SY91,AY94,ASY00}. 
It would be interesting to 
construct 
the 
corresponding classical $r$-matrix 
a la 
Sklyanin~\cite{Skly82,Semenov}. 

\section{Space-discrete 
massive Thirring model 
}
\label{sec3}

In this section, 
we 
describe how to 
discretize one of the two independent variables 
of the massive Thirring model 
in light-cone (or characteristic) coordinates~\cite{KN2,Morris79,GIK}, 
which is 
the first negative flow 
of the Chen--Lee--Liu hierarchy~\cite{NCQL,Linden1}
(see~\cite{Kuz,KaMoIno} for the model in 
laboratory coordinates). 
With a slight abuse of terminology, 
we will call it a 
``space-discretization" of the  
massive Thirring model, 
although it 
would 
be more natural and 
appropriate to use it 
for the massive Thirring model 
in laboratory coordinates.

The generalized binary 
B\"acklund--Darboux transformation (\ref{ALDB1}) 
with suitably defined intermediate potentials $u_m$ and $v_m$ 
can preserve 
the Ablowitz--Ladik 
spectral problem (\ref{AL-Ln}) form-invariant, 
as well as an infinite set of 
isospectral time-evolution equations 
such as 
(\ref{AL-Mn1}) and (\ref{AL-Mn2}). 
Thus, 
the temporal Lax matrix $M_n$ 
given 
by 
(\ref{Lax-Mn1}) with (\ref{FG_choice}) 
is not the 
only possible choice; 
it can be replaced with 
any $M_n$-matrix 
corresponding to a flow of the Ablowitz--Ladik hierarchy, 
or equivalently, 
the continuous Chen--Lee--Liu hierarchy. 
To obtain a space-discrete analog of the massive Thirring model,  
%
we 
suppress 
the lattice index $m$ 
for the Ablowitz--Ladik hierarchy and 
consider the isospectral time-evolution equation (\ref{AL-Mn2}) 
written 
in the 
form: 
%
\begin{align}
& \left[
\begin{array}{c}
 \Psi_{1,n} \\
 \Psi_{2,n} \\
\end{array}
\right]_{t
}
= 
\left[
\begin{array}{cc}
 O & \frac{1}{\zeta} Q_n \\
 \frac{1}{\zeta} R_n & 
 \frac{1}{\zeta^2} I - R_n Q_n \\
\end{array}
\right]
\left[
\begin{array}{c}
 \Psi_{1,n} \\
 \Psi_{2,n} \\
\end{array}
\right].
\nonumber
\end{align}
Then, 
substituting the Lax pair
given by 
(\ref{Lax-Ln1}) and 
\begin{align}
 M_n = \left[
\begin{array}{cc}
 O & \frac{1}{\zeta} Q_n \\
 \frac{1}{\zeta} R_n & 
 \frac{1}{\zeta^2} I - R_n Q_n \\
\end{array}
\right] 
\nonumber
\end{align}
into 
the zero-curvature equation (\ref{Lax_eq}), we obtain 
a set of six 
equations. 
A direct calculation shows that 
only four of them are independent, 
which can be 
presented 
as 
\begin{equation} 
\label{sdMTM1}
\left\{ 
\begin{split}
& \beta \delta \left( \gamma Q_{n+1} - \alpha Q_n \right) 
+ \left( \alpha \beta - \gamma \delta \right) u_n -u_n v_n 
\left( \beta Q_{n+1} - \delta Q_n \right) = O, 
\\[1mm]
& \beta \delta \left( \alpha R_{n+1} - \gamma R_n \right) 
- \left( \alpha \beta - \gamma \delta \right) v_n -
\left(  \delta R_{n+1} - \beta R_n \right) u_n v_n  = O,
\\[1mm]
& \left( \alpha \beta - \gamma \delta \right) u_{n,t} 
 + \alpha \gamma \left(\beta Q_{n+1} - \delta Q_n \right)
  + u_n \left( \alpha R_{n+1} - \gamma R_n \right) u_n 
\\ & \hspace{10mm}
  - u_n \left(\alpha \beta R_{n+1} Q_{n+1} - \gamma \delta R_n Q_n \right) =O, 
\\[1mm]
& \left( \alpha \beta - \gamma \delta \right) v_{n,t} 
 + \alpha \gamma \left( \delta R_{n+1} -  \beta R_n \right)
  + v_n \left(  \gamma Q_{n+1} - \alpha Q_n \right) v_n 
\\ 
& \hspace{10mm}
  - \left( \gamma \delta R_{n+1} Q_{n+1} - \alpha \beta R_n Q_n \right) v_n =O.
\end{split}
\right. 
\end{equation}
By setting \mbox{$\gamma=\alpha=\mathrm{i}$} 
and 
\mbox{$\delta=-\beta=-2\mathrm{i}/\varDelta 
$} 
where \mbox{$\varDelta \in \mathbb{R}$} 
is 
a lattice parameter, 
(\ref{sdMTM1})
reads 
\begin{equation} 
\label{sdMTM2}
\left\{ 
\begin{split}
&  \frac{2}{\varDelta} \left( Q_{n+1} - Q_n \right) 
+ 2 \mathrm{i} u_n - u_n v_n \left(  Q_{n+1} + Q_n \right) = O, 
\\[1mm]
&  \frac{2}{\varDelta} \left( R_{n+1} - R_n \right) 
 - 2 \mathrm{i} v_n 
 + \left( R_{n+1} + R_n \right) u_n v_n  = O,
\\[1mm]
&  2 u_{n,t} 
 + \mathrm{i} \left( Q_{n+1} + Q_n \right)
  - \frac{\mathrm{i}\varDelta}{2} 
 u_n \left( R_{n+1} -  R_n \right) u_n 
\\ 
& \hspace{10mm}
  - u_n \left(R_{n+1} Q_{n+1} + R_n Q_n \right) =O, 
\\[1mm]
& 2 v_{n,t} 
 - \mathrm{i} \left(  R_{n+1} + R_n \right)
  - \frac{\mathrm{i}\varDelta}{2} 
 v_n \left( Q_{n+1} - Q_n \right) v_n 
\\ 
& \hspace{10mm}
  + \left( R_{n+1} Q_{n+1} + R_n Q_n \right) v_n =O.
\end{split}
\right. 
\end{equation}
Note that 
the set of coefficients 
in (\ref{sdMTM2})
can be changed by 
rescaling the variables. 
System (\ref{sdMTM2}) admits the Hermitian conjugation reduction 
\mbox{$R_n = \mathrm{i} Q_n^\dagger$}, 
\mbox{$v_n = \mathrm{i} u_n^\dagger$}, 
so 
we obtain 
a proper 
space-discretization of 
the massive Thirring model as 
\begin{equation} 
\label{MTM1}
\left\{ 
\begin{split}
&  \frac{2}{\varDelta} \left( Q_{n+1} - Q_n \right) 
+ 2 \mathrm{i} u_n - \mathrm{i} u_n u_n^\dagger \left(  Q_{n+1} + Q_n \right) = O,
\\[1mm]
&  2 u_{n,t} 
 + \mathrm{i} \left( Q_{n+1} + Q_n \right)
 + \frac{\varDelta}{2} \hspace{1pt} u_n \left( Q_{n+1}^\dagger - Q_n^\dagger \right) u_n 
\\ 
& \hspace{10mm}
  - \mathrm{i} u_n \left(Q_{n+1}^\dagger Q_{n+1} + Q_n^\dagger Q_n \right) =O.
\end{split}
\right. 
\end{equation}
In the continuous space 
limit, 
(\ref{MTM1}) indeed reduces 
to the massive Thirring model 
in light-cone 
coordinates (see (3.48) in~\cite{Linden1}): 
\begin{equation} 
\nonumber
\left\{ 
\begin{split}
&  Q_x + \mathrm{i} u - \mathrm{i} u u^\dagger Q = O,
\\[0.5mm]
&  u_{t} + \mathrm{i} Q
  - \mathrm{i} u Q^\dagger Q =O.
\end{split}
\right. 
\end{equation}
We remark that 
another 
lattice 
version 
of  
the massive Thirring model in light-cone coordinates 
was studied 
in~\cite{NCQ83}. 
%
%
%


\section{Concluding remarks}


In this paper, we have developed an effective 
approach for 
generating 
new integrable lattice systems 
from 
B\"acklund--Darboux transformations for known integrable systems. 
The idea to 
interpret 
a 
B\"acklund--Darboux transformation 
as a discrete spectral problem 
is already well-known (see, {\it e.g.}, \cite{Chud1,Chud2,DJM83}); 
the main new feature of 
our 
approach
is to 
consider
the intermediate potentials appearing in a binary 
B\"acklund--Darboux transformation as new dependent variables 
in the discrete spectral problem, up to a rescaling of the variables. 
As the name implies, 
the binary B\"acklund--Darboux transformation 
is equivalent to 
the composition of 
two elementary 
B\"acklund--Darboux transformations 
in 
either 
order 
of the 
composition~\cite{Kono79,Kono82,Konop3}. 
In the example of the Ablowitz--Ladik hierarchy, 
$u_m$ and $v_m$ 
in (\ref{ALDB1})
appear 
in the 
decomposition of the binary B\"acklund--Darboux transformation
into 
two permutable 
elementary 
B\"acklund--Darboux transformations 
as~\cite{Rourke}
\begin{equation} 
(Q_m,R_m) 
\rightarrow (u_m, \ast) 
\rightarrow (\widetilde{Q}_m, \widetilde{R}_m)
\nonumber 
\end{equation}
and 
\begin{equation}
(Q_m,R_m) 
\rightarrow (\ast, v_m) 
\rightarrow (\widetilde{Q}_m, \widetilde{R}_m)
\nonumber
\end{equation}
up to a rescaling and 
a 
space translation, 
so they are indeed the intermediate potentials; 
a similar 
decomposition 
holds true for 
the continuous 
Chen--Lee--Liu hierarchy 
if one considers the pair of variables \mbox{$(Q_m,R_{m-1})$} 
or \mbox{$(Q_{m-1},R_{m})$}  
(cf.~(\ref{AL-Mn1}) or (\ref{AL-Mn2})). 
Then, we 
associate the new discrete spectral problem 
with a suitable isospectral 
time-evolution equation on a case-by case consideration 
and 
obtain 
an evolutionary lattice system 
from the compatibility condition called 
the zero-curvature  
equation. 
We illustrated this 
approach 
by deriving the general lattice system 
(\ref{sdCLL2}), which involves the arbitrary parameters 
$\alpha$, $\beta$, $\gamma$ and $\delta$
and includes 
the space-discrete Chen--Lee--Liu system (\ref{sdCLL3}) as a special case; 
the derivation of 
negative flows 
of the 
integrable hierarchy 
is easier and can be performed in a 
more 
systematic manner 
as is illustrated 
in section~\ref{sec3}. 
This approach 
is 
quite 
useful for obtaining 
proper discretizations of 
integrable systems 
that admit 
the complex/Hermitian 
conjugation reduction 
between the two dependent variables, 
such as 
the NLS 
system, 
derivative NLS 
systems and their matrix generalizations; 
this is because the 
intermediate potentials 
generally 
inherit 
the 
internal symmetries 
of 
the original 
continuous system, 
which guarantees the feasibility of such a 
reduction. 

In the continuous case, 
by applying a nonlocal 
transformation of 
dependent 
variables~\cite{Kun,WS,NCQL,KN
}, 
we can 
transform 
the Chen--Lee--Liu 
equation
to 
other derivative NLS 
equations 
such as 
the Kaup--Newell 
equation~\cite{KN}
and the Ablowitz--Ramani--Segur (Gerdjikov--Ivanov) 
equation~\cite{ARS,GI}; 
the same transformation applies to other flows of the integrable 
hierarchy 
including 
the massive Thirring 
model~\cite{KN2,GIK,NCQL,
DHS84} (also see \cite{David84}). 
Note that the nonlocal 
quantity used in 
this transformation 
is 
given by 
the first (or second) component of the linear eigenfunction 
of the Lax-pair representation 
with the spectral parameter set equal to 
zero. 
In a similar manner, 
we 
can 
obtain 
new 
proper space-discretizations of 
other derivative 
NLS systems 
from the space-discrete Chen--Lee--Liu 
system (\ref{sdCLL3}) 
by applying a discrete analog of the nonlocal transformation. 
Indeed, (\ref{gDB2}) with 
(\ref{2nd_exp}) implies 
that 
the quantity \mbox{$Z_n := (-\delta)^{-n} \mathrm{e}^{\xi t} 
\lim_{\zeta \to 0}   \zeta^{n} \Psi_{1,n} 
$} 
satisfies the following relation 
for any constant $\xi$:  
\begin{align}
Z_{n+1} = 
\left( \alpha \beta  I - u_n v_n \right) 
\left( -\gamma \delta I + u_n v_n \right)^{-1} Z_n. 
\nonumber
\end{align}
The 
time derivative of $Z_n$ for a suitably chosen $\xi$ 
can be 
obtained from (\ref{general_M}) with 
(\ref{Lax-Mn1}) 
and (\ref{FG_choice}) as 
\begin{align}
Z_{n,t} = \left( \alpha \beta - \gamma \delta \right)
\left[ a \beta \gamma \left( \beta \gamma I - u_{n-1} v_n \right)^{-1}  
+ b \alpha \delta \left( \alpha \delta I - u_{n} v_{n-1} \right)^{-1}  \right] Z_n. 
\nonumber
\end{align}
Thus, 
the nonlocal transformation
\[
q_n := Z_n^{-1} u_n, \hspace{5mm} r_n := v_n Z_n
\]
or its one-parameter generalization in the scalar case
\[
q_n := Z_n^{-k} u_n, \hspace{5mm} r_n := v_n Z_n^{k}
\]
can be applied 
to (\ref{sdCLL2}); 
by setting \mbox{$\gamma=\alpha$},  
\mbox{$\delta=-\beta$} and \mbox{$\alpha \beta=1$}, 
(\ref{sdCLL2}) reduces to 
(\ref{sdCLL3}) and 
we obtain proper space-discretizations of other derivative NLS systems. 

%

In the original formulation of 
B\"acklund--Darboux transformations~\cite{Kono82,Chud1,Chud2,Sall82,MaSa91,Pemp}, 
the intermediate potentials
such as $u_m$ and $v_m$ in (\ref{ALDB1}) 
can be 
written 
explicitly 
in terms of the linear eigenfunctions  
at 
some fixed 
values of the spectral parameter 
$\zeta$. 
Then, 
B\"acklund--Darboux transformations 
can be applied 
iteratively 
to 
obtain 
a sequence of new solutions of 
a nonlinear 
integrable system 
from 
its seed 
solution and the associated linear eigenfunctions; 
the final result does not depend on the order of 
applications. 
That is, 
B\"acklund--Darboux transformations with (generally) different values of 
the B\"acklund parameters 
are mutually 
commutative 
as long as boundary conditions are fixed appropriately; 
this fact can 
be understood intuitively 
by identifying 
B\"acklund--Darboux transformations
as 
discrete-time flows 
that belong to 
the same integrable hierarchy. 
Without taking into account 
how to 
express the intermediate potentials 
in terms of the linear eigenfunctions of the Lax pair, 
the permutability 
condition for B\"acklund--Darboux transformations 
results in a matrix re-factorization problem 
(see, {\it e.g.}, \cite{Ve03,GoVe04,Ve07}). 
In the case considered in this paper, 
it reads~\cite{Chud2,GoVe04,Ve07,SuVe03,BoSu08}
\begin{align}
& L (\mathcal{U}', \mathcal{V}'; \mu, \nu, \xi, \eta)
L (u, v; \alpha, \beta, \gamma, \delta) 
= L (u', v'; \alpha, \beta, \gamma, \delta) 
L (\mathcal{U}, \mathcal{V}; \mu, \nu, \xi, \eta), 
\label{YB_Lax}
\end{align}
where 
the 
matrix 
$L (u, v; \alpha, \beta, \gamma, \delta)$ 
is given by (\ref{Lax-Ln1}) with the unnecessary lattice index $n$ removed 
and the 
dependence 
on the spectral parameter $\zeta$ 
suppressed. 
Considering the matrix inverse 
of $L$, 
we can rewrite (\ref{YB_Lax}) as~\cite{SuVe03,BoSu08}
\begin{align}
& L (u, v; \gamma, \delta, \alpha, \beta) 
L (\mathcal{U}', \mathcal{V}'; \xi, \eta, \mu, \nu)
= L (\mathcal{U}, \mathcal{V}; \xi, \eta, \mu, \nu) 
L (u', v'; \gamma, \delta, \alpha, \beta). 
\label{YB_Lax2}
\end{align}
Equation (\ref{YB_Lax}) or (\ref{YB_Lax2}) 
can be naturally 
represented 
using an elementary quadrilateral 
and 
defines 
a parameter-dependent Yang--Baxter map 
\[
(u, v; \mathcal{U}', \mathcal{V}') 
\mapsto 
(u', v'; \mathcal{U}, \mathcal{V}),
\]
wherein 
the 
fields 
are assigned to 
the edges 
of 
the quadrilateral, 
instead of the vertices~\cite{BoSu08,SuVe03,Ve07,PaToVe}. 
The Lax
(or zero-curvature) 
representation for the Yang--Baxter map 
is given by 
(\ref{YB_Lax2})~\cite{GoVe04,SuVe03,Ve07,BoSu08}. 
%
%
It is not evident from the explicit form of the Lax 
matrix in (\ref{Lax-Ln1}) 
that 
(\ref{YB_Lax2})
can be solved uniquely 
to provide 
the Yang--Baxter map, 
but the 
map 
itself 
can, in principle, 
be 
constructed using 
the definition of the intermediate potentials $u$ and $v$ 
in terms of the linear eigenfunctions. 


\end{document}